\documentclass[journal]{IEEEtran}
\usepackage{cite}
\usepackage{graphicx}
\usepackage{subfigure}
\usepackage{amsmath}
\usepackage{dcolumn}
\usepackage{bm}

\begin{document}

\title{Comparison of coherence times \\ in three dc SQUID phase
qubits}

\author{Hanhee Paik, B. K. Cooper, S. K. Dutta, R. M. Lewis, R. C. Ramos, T. A. Palomaki, A. J. Przybysz,
\\ A. J. Dragt, J. R. Anderson, C. J. Lobb, and F. C. Wellstood
\thanks{Manuscript received August 30, 2006. This work was supported by NSF under
the Qubic program, the Laboratory for Physical Sciences, College
Park, MD, and the Center for Superconductivity Research at the
University of Maryland, College Park, MD 20742 USA.  Hanhee Paik, B.
K. Cooper, S. K. Dutta, R. M. Lewis, R. C. Ramos, T. A. Palomaki, A.
J. Przybysz, A. J. Dragt, J. R. Anderson, C. J. Lobb, and F. C.
Wellstood are with the the University of Maryland, College Park, MD
20742 USA. Corresponding author email: hanhee@umd.edu}}
\markboth{3EB03}{Shell \MakeLowercase{\textit{et al.}}: Journal of
Applied Superconductivity 2004}

\maketitle

\begin{abstract}
We report measurements of spectroscopic linewidth and Rabi
oscillations in three thin-film dc SQUID phase qubits. One device
had a 6-turn Nb loop, the second had a single-turn Al loop, and the
third was a first order gradiometer formed from 6-turn wound and
counter-wound Nb coils to provide isolation from spatially uniform
flux noise. In the 6 - 7.2 GHz range, the spectroscopic coherence
times for the gradiometer varied from 4 ns to 8 ns, about the same
as for the other devices (4 to 10 ns). The time constant for decay
of Rabi oscillations was significantly longer in the single-turn Al
device (20 to 30 ns) than either of the Nb devices (10 to 15 ns).
These results imply that spatially uniform flux noise is not the
main source of decoherence or inhomogenous broadening in these
devices.
\end{abstract}


\IEEEpeerreviewmaketitle

\section{Introduction}

\PARstart{D}ESPITE much recent progress in the use of
superconducting devices for quantum computation \cite{nori},
decoherence still presents a major challenge.  For Josephson phase
qubits \cite{han, jm1, ajb, skd, bui, jm2, jm3}, Martinis \textit{et
al.} \cite{jm4} have proposed that dielectric loss and two-level
fluctuators in dielectrics are the primary cause of decoherence.
They showed significant improvement could be obtained by replacing
lossy dielectrics with lower-loss materials.  Van Harlingen
\textit{et al.} \cite{harl1} argued that while critical current
fluctuations would produce decoherence, the observed coherence times
in flux and phase qubits were much shorter than would be expected
from the level of critical current noise that has been typically
observed in tunnel junctions \cite{fcw1, harl2, fcw2}. Similarly,
Martinis \textit{et al.} \cite{jm5} have argued that charge noise
should have a small impact on the coherence time of phase qubits due
to their large junction capacitance.  Recently, Bertet \textit{et
al.} reported that decoherence in their flux qubit came from the
detection dc SQUID \cite{mooij}.

Flux noise is another possible source of decoherence in phase qubits
as most are essentially rf or dc SQUIDs. In this paper, we compare
Rabi oscillations and spectroscopic coherence times in three dc
SQUID phase qubits that were built with a single-turn magnetometer
configuration (AL1), a multi-turn magnetometer configuration (NB1)
and a gradiometer configuration (NBG) respectively. Although we did
not perform a direct test on the gradiometer balance, the
counter-wound configuration should make it much less sensitive to
spatially unform magnetic fields than either of the magnetometers.

\begin{figure}
\centering
\includegraphics[width=3 in]{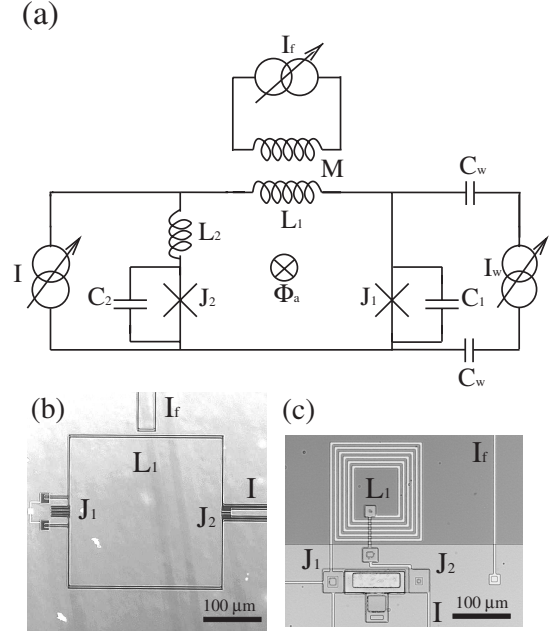}
\caption{(a) Schematic of dc SQUID phase qubit. I is the current
bias, I$_f$ is current for the flux bias, M is mutual inductance
between the flux bias coil and the SQUID loop and $\Phi_a$ is the
applied flux in the SQUID loop. C$_1$ and C$_2$ are the capacitances
of the qubit junction J$_1$ and the isolation junction J$_2$,
respectively. Microwave source I$_w$ is coupled to J$_1$ through
capacitors C$_w$. Photographs of (b) Al single-turn SQUID
magnetometer AL1 and (c) Nb 6-turn SQUID magnetometer NB1.
}\label{fig1}
\end{figure}
\section{dc SQUID phase qubits without and with gradiometer loops}
Figure 1(a) shows a schematic of a dc SQUID phase qubit \cite{jm1}.
We refer to J$_1$ as the qubit junction and J$_2$ as the isolation
junction. In this qubit design, junction J$_2$ is needed to read out
the state of J$_1$ via tunneling to the voltage state \cite{jm1}.
J$_2$ and inductor L$_1$ also inductively isolate the qubit from
current noise on the bias leads. By choosing L$_1$ $\gg$ L$_2$ +
L$_{j2}$, where L$_{j2}$ is the Josephson inductance of the
isolation junction, current noise will be mainly diverted through
J$_2$ rather than the qubit junction J$_1$.

Figures 1(b) and 1(c) show two of our SQUID phase qubits that have
magnetometer loops.  Device NB1 [see Fig. 1(b)] is a thin-film Nb
magnetometer with a 6-turn loop. The device was made at Hypres,
Inc., from a Nb$/$AlO$_x/$Nb trilayer using their 100 A/cm$^2$
process. The qubit junction has an area of 100 $\mu$m$^2$ and we
applied a small magnetic field in the plane of the junctions to
reduce the critical current of the device to the 10 to 30 $\mu$A
range. Subsequent measurements on similar devices, which were not
suppressed by magnetic field, yielded similar spectroscopic
coherence times and Rabi decay times \cite{skdthesis}.

Device ALl is a single-turn dc SQUID magnetometer made from
thin-film Al [see Fig. 1(c)]. We used photolithography and
double-angle evaporation to form the loop and the Al$/$AlO$_x/$Al
tunnel junctions.  The qubit junction has an area of the 80 $\mu
m^2$. Other than AlO$_x$, no insulation layers were deposited on
this device.

\begin{table}
\centering \caption{\label{tab:table1} Parameters for SQUIDs NBG,
NB1 and AL1.  I$_{01}$ and I$_{02}$ are the critical currents of
J$_1$ and J$_2$, respectively.  L$_{j1}$(0) and L$_{j2}$(0) are the
Josephson inductances of J$_1$ and J$_2$ when they are unbiased and
$\beta = (I_ {01}+I_{02})L/\Phi_0$, where L = L$_1$ + L$_2$ is the
total inductance of the SQUID loop.}
\begin{tabular}{cccc}
\hline\\[3pt]
& gradiometer & magnetometer & magnetometer \\
Parameters & NBG & NB1 & AL1 \\[3pt]
\hline\\[3pt]
I$_{01}$ ($\mu$A) & 23.0 & 33.8 & 21.275  \\[3pt]
I$_{02}$ ($\mu$A) & 3.8 & 4.8 & 9.445  \\[3pt]
C$_{1}$ (pF) & 4.1 & 4.4 & 4.1 \\[3pt]
C$_{2}$ (pF) & 2.0 & 2.2 & 2.1 \\[3pt]
L$_{j1}(0)$ (pH) & 13.9 & 9.7 & 13.2 \\[3pt]
L$_{j2}(0)$ (pH) & 84.9 & 68 & 44.5 \\[3pt]
L$_{1}$ (pH) & 4540 & 3530 & 1236 \\[3pt]
L$_{2}$ (pH) & 12 & 20 & 5 \\[3pt]
$\beta$ & 34 & 66 & 19 \\[3pt]
\hline
\end{tabular}
\footnotetext[1]{suppressed} \footnotetext[2]{unsuppressed}
\end{table}

Device NBG (see Fig. 2) was made at Hypres from a Nb$/$AlO$_x/$Nb
trilayer using their 30 A/cm$^2$ process. The qubit junction has an
area of 102 $\mu$m$^2$.  The SQUID has two 6-turn thin-film Nb coils
wound in opposition to form a magnetic field gradiometer.  To apply
a net flux to the device, we placed a flux bias line on the right
side of the device (closer to coil L$_{1a}$ than to L$_{1b}$ in Fig.
2). All three devices were made on silicon wafers with a layer of
thermally grown silicon dioxide.

\begin{figure}
\centering
\includegraphics[width= 3 in]{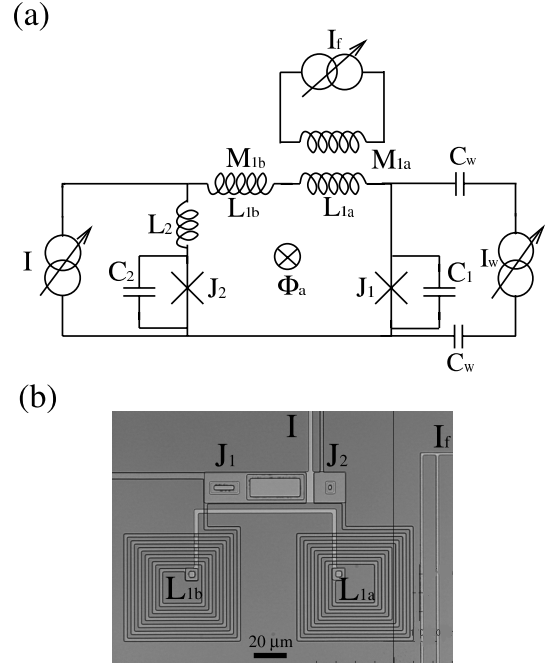}
\caption{(a) Schematic of dc SQUID phase qubit with a gradiometer. I
is the current bias, I$_f$ is current for the flux bias, $\Phi_a$ is
the applied flux in the SQUID loop. C$_1$ and C$_2$ are the
capacitances of junctions J$_1$ and J$_2$.  L$_{1a}$ and L$_{1b}$
are inductances of each coil of the gradiometer, M$_{1a}$ and
M$_{1b}$ are mutual inductances between each coil, respectively, and
the flux bias line. A microwave source I$_w$ is coupled to J$_1$
through capacitors C$_w$. (b) Photo of dc SQUID phase qubit,
gradiometer NBG.} \label{fig2}
\end{figure}

\section{Experimental setup}
Device NB1 was measured on an Oxford Instruments Kelvinox model 200
dilution refrigerator at a base temperature of 25 mK, while devices
AL1 and NBG were measured on an Oxford Instruments model 25 dilution
refrigerator at a base temperature of 80 to 100 mK. Each device was
mounted in a closed superconducting aluminum box to shield out
magnetic fields. In addition, each refrigerator was surrounded by a
room-temperature mu-metal shield and enclosed in an rf-shielded
room. To characterize the devices, we measured the switching current
as a function of the applied flux. By fitting the characteristics to
those of an ideal dc SQUID we obtained the device parameters,
summarized in Table 1.

\section{Measurement of energy levels}
Before making any measurements on a SQUID, we use a flux shaking
technique to initialize the flux state \cite{tauno}. We then apply a
simultaneous flux and current ramp so as to bias the qubit junction
with current, and not the isolation junction. With this biasing
scheme, our device acts as an ideal phase qubit with the two lowest
levels in a well of the tilted washboard potential forming the qubit
states, $|0\rangle $ and $|1\rangle$.

The qubit state can be monitored by measuring the rate at which the
system tunnels to the voltage state \cite{jm1}; the first excited
state $|1\rangle$ typically tunnels about 500 times faster than the
ground state $|0\rangle $. During the simultaneous current and flux
ramp, we record the time at which the device escapes to the voltage
state. We repeat this process on the order of 10$^5$ times to build
up a histogram of escape events versus ramp time, which we convert
to escape rate versus current (for spectroscopy) or escape rate
versus time (for Rabi oscillations).

\begin{figure}
\centering
\includegraphics[width=3 in]{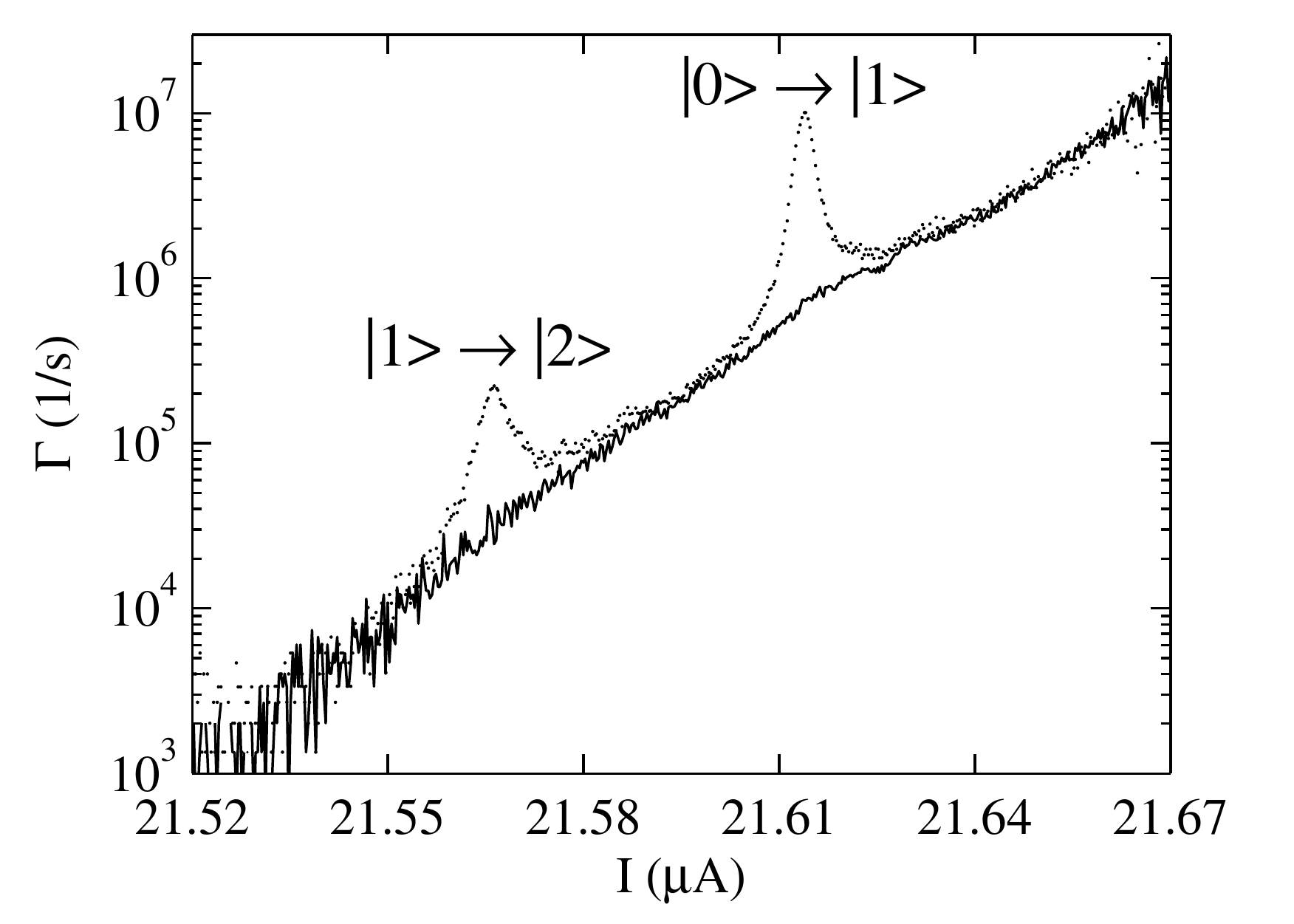}
\caption{Total escape rate vs. current for NBG at 100 mK.  Dotted
line is when 6.6 GHz microwave is applied to the qubit junction and
solid line is without microwaves.}\label{G}
\end{figure}

As an example, Fig. \ref{G} shows the total escape rate of the qubit
junction in device NBG as a function of the bias current I, with and
without application of 6.6 GHz microwaves.  Sweeping current through
the qubit changes the energy level spacing adiabatically.  When the
microwaves comes into resonance with the energy level spacing,
transitions to the excited state occur and we see enhancement in the
total escape rate.  In Fig \ref{G}, two clear resonance peaks are
visible, at around 21.57 $\mu$A and 21.62 $\mu$A, corresponding to
$|1\rangle \rightarrow |2\rangle$ and $|0\rangle \rightarrow
|1\rangle$ transitions.

\begin{figure}
\centering
\includegraphics[width=3in]{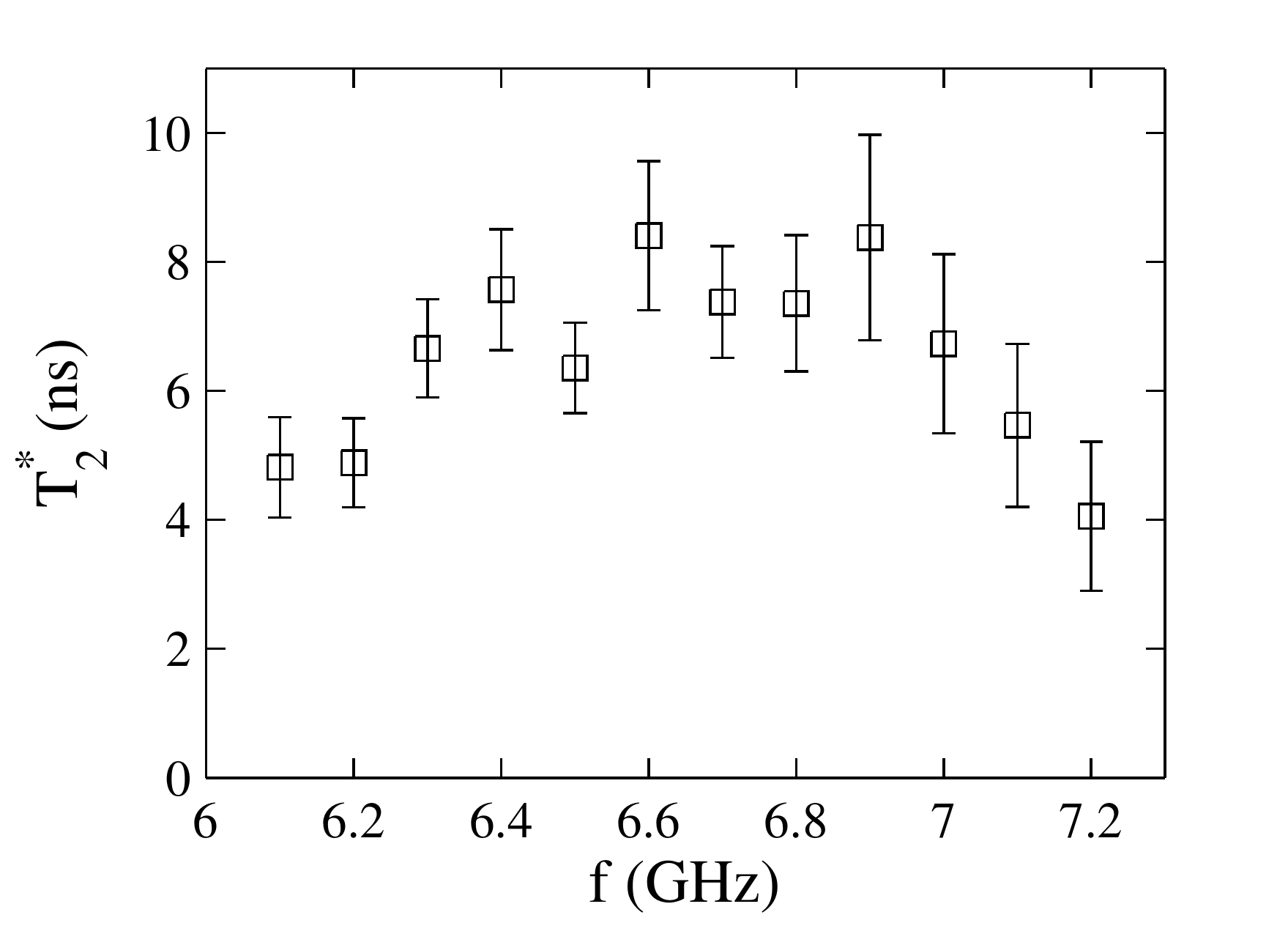}
\caption{Spectroscopic coherence time T$_2^*$ of the $|0\rangle$ to
$|1\rangle$ transition versus frequency for SQUID gradiometer NBG
measured at 100 mK.\label{t2star2}}
\end{figure}
\section{measurement of T$_2^*$}

To determine the spectroscopic coherence time T$_2^*$ \cite{xu} of
the $|0\rangle \rightarrow |1\rangle$ transition frequency f$_{01}$,
we obtained the low-power half-width at half-maximum $\Delta$I of
the resonance peak and recorded its location I(f$_{01}$). Repeating
this procedure for a range of applied microwave frequencies yields
f$_{01}$ and $\Delta$I as a function of the ramp current I. The
spectroscopic coherence time as a function of the frequency was then
found from \cite{xu}
\begin{equation}
T_2^* = \frac{1}{2\pi\Delta I \frac{\partial f_{01}}{\partial I}}.
\end{equation}

Figure \ref{t2star2} shows a plot of T$_2^*$ versus microwave
frequency for gradiometer NBG, measured at 100 mK. For frequencies
in the 6.0 - 7.2 GHz range, T$_2^*$ varied between about 4 ns and 8
ns. Spectroscopic measurements on magnetometers NB1 and AL1 revealed
comparable variations in T$_2^*$, from about 4 to 10 ns in the same
frequency range \cite{skdthesis}. Since T$_2^*$ is sensitive to
low-frequency noise (inhomogenous broadening), as well as pure
dephasing and dissipation \cite{xu}, we can conclude that the
combined effect of low-frequency noise, pure dephasing and
dissipation is comparable in the three devices.

\section{Comparisons of Rabi oscillations between a gradiometer and
magnetometers}
\begin{figure}
\centering
\includegraphics[width=3 in]{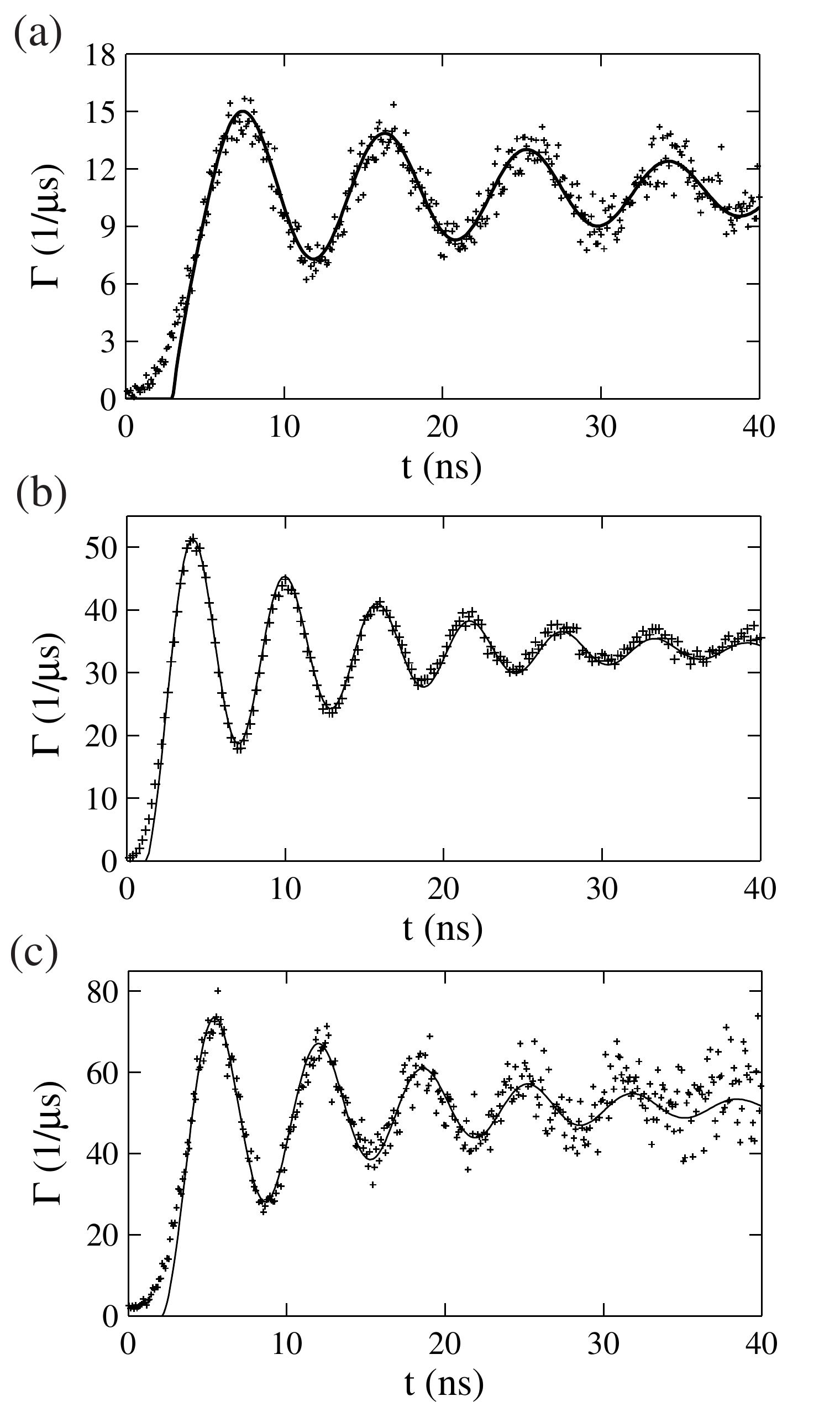}
\caption{Measurements of Rabi oscillation in the escape rate in (a)
single-turn magnetometer AL1, (b) 6-turn magnetometer NB1, (c)
gradiometer NBG. In each case the solid curve is a least-square fit
to a cosine function with an exponential decay envelope.
\label{rabi}}
\end{figure}
To distinguish the effects of low-frequency noise from dephasing
processes, we also measured Rabi oscillations. The idea is that the
envelope decay time constant T$'$ of a Rabi oscillation is sensitive
to noise at the Rabi frequency while the main effect of noise at
much lower frequencies (which acts like inhomogenous broadening) is
to change the shape of the envelope \cite{jm5, harl1}.  This
relative insensitivity  of the Rabi oscillation to low-frequency
noise is similar to the situation in a spin-echo measurement
\cite{jm5}. The envelope decay time constant T$'$, the energy
relaxation time T$_1$, and the coherence time T$_2$ are related by
\cite{torrey}
\begin{equation}
\frac{1}{T'} = \frac{1}{2T_1} + \frac{1}{2T_2},
\end{equation}
when the Rabi oscillation is driven on resonance. Although spin-echo
measurements are the best way to directly determine T$_2$ and
distinguish pure dephasing from inhomogenous broadening, we were not
able to measure spin echoes in these devices due to their relatively
short coherence times. Finally, from seperate measurements, we find
T$_1$ is roughly 50 ns.

Figure 4 shows typical examples of measured Rabi oscillations in the
total escape rate for the three devices.  We applied microwave
frequencies of 7.6 GHz for NB1 at 25 mK [see Fig. 4(a)], 7 GHz for
AL1 at 80 mK [see Fig. 4(b)] and 6.5 GHz for NBG at 100 mK [see Fig.
4(c)].  The applied microwaves coupled to the qubit junction via a
small capacitor [See Fig. 1 and Fig. 2] and resonantly drove the
qubit between $|0\rangle$ and $|1\rangle$. In each case, we observed
clear oscillations in the escape rate.  In these plots, t = 0
indicates when the microwaves were turned on.

The solid curves in Fig. 4 are least-square fits to a cosine
function with an exponential envelope with time constant T$'$.  From
the fits, T$'$ for the gradiometer NBG was about 12 ns, while for
magnetometers NB1 and AL2 it was about 12 ns and 27 ns, respectively
(see Table 2).  Density matrix simulations reveal that the escape
rate we observe is dominated by a small population in $|2\rangle$
that escapes very rapidly ($\Gamma_2 \sim 10^{10}$ 1/s $\gg$
1/T$'$), and this population is directly proportional to the
occupancy of $|1\rangle$ which escapes much more slowly ($\Gamma_1
\sim 10^7$ 1/s). While tunneling contributes to spectroscopic
broadening \cite{xu}, measurements over a wide range of conditions
with different escape rates did not alter T$'$ for Rabi oscillations
in the total escape rate \cite{skdthesis}.

Thus the single-turn aluminum magnetometer AL1 had a substantially
longer envelope decay time T$'$ than either the Nb magnetometer or
the Nb gradiometer. From this comparison, we can safely conclude
that T$'$ in our dc SQUID phase qubits is not being limited by
spatially uniform flux noise.  Finally, we note that since T$' \sim$
2T$_2^*$ (see Table 2), low-frequency noise is not causing
significant inhomogeneous broadening of the resonance.

\begin{table}
\centering \caption{\label{tab:table2}Summary of spectroscopic
coherence time T$_2^*$ and time constant T$'$ for decay of Rabi
oscillation.}
\begin{tabular}[c]{cccc}
\hline \\
 & gradiometer & magnetometer & magnetometer\\
 & NBG & NB1 & AL1 \\[3pt]
\hline \\
T$'$(ns) & 10 - 15 & 10 - 15 & 20 - 30 \\[3pt]
T$_2^*$(ns) & 4 - 8 & 4 - 10 & 4 - 10 \\[3pt]
\hline \\
\end{tabular}
\end{table}

\section{Conclusion}
In conclusion, we have measured the spectroscopic coherence time
T$_2^*$ and Rabi oscillations in three dc SQUID phase qubits. One
device was a Nb gradiometer with 6-turn wound and counter-wound
coils, the second was an Al magnetometer made with a single-turn
loop, and the third was a Nb magnetometer with a 6-turn coil. The
gradiometer did not show significantly longer T$'$ or T$_2^*$, and
in fact the single-turn Al magnetometer showed a significantly
longer T$'$ than either the Nb gradiometer or Nb magnetometer.  We
conclude that spatially uniform flux noise is not a dominant source
of decoherence in our phase qubits. We cannot rule out the
possibility of a local source of flux noise, which would not be
nulled by a gradiometer.

\section*{Acknowledgment}
The authors would like to thank M. Mandelberg, M. Manheimer, B.
Palmer and F. W. Strauch for useful discussions.

\end{document}